\documentclass[twocolumn,superscriptaddress,nofootinbib,notitlepage,longbibliography,aps]{revtex4-1}
\pdfoutput=1

\usepackage{graphicx}
\usepackage{amsmath}
\usepackage{amssymb}
\usepackage{amsfonts}
\usepackage[dvipsnames]{xcolor}
\usepackage[linktoc=none]{hyperref}
\usepackage[utf8]{inputenc}
\usepackage{comment}

\hypersetup{colorlinks=true,linkcolor=blue,citecolor=teal,filecolor=magenta, urlcolor=cyan}

\newcommand{\UNINA}{Dipartimento di Fisica ``Ettore Pancini'', Università degli studi di Napoli ``Federico II'', Complesso Universitario Monte S. Angelo, I-80126 Napoli, Italy}
\newcommand{\INFN}{INFN - Sezione di Napoli, Complesso Universitario Monte S. Angelo, I-80126 Napoli, Italy}
\newcommand{\SSM}{Scuola Superiore Meridionale, Università degli studi di Napoli ``Federico II'', Largo San Marcellino 10, 80138 Napoli, Italy}

\begin{document}

\title{Light burden of memory: \\Constraining primordial black holes with high-energy neutrinos}
\author{Marco Chianese}
\email{marco.chianese@unina.it}
\affiliation{\UNINA}
\affiliation{\INFN}
\author{Andrea Boccia}
\affiliation{\SSM}
\affiliation{\INFN}
\author{Fabio Iocco}
\affiliation{\UNINA}
\affiliation{\INFN}
\author{Gennaro Miele}
\affiliation{\UNINA}
\affiliation{\INFN}
\affiliation{\SSM}
\author{Ninetta Saviano}
\affiliation{\INFN}
\affiliation{\SSM}

\begin{abstract}
Recent studies point out that quantum effects, referred to as ``memory burden'', may slow down the evaporation of black holes. As a result, a population of light primordial black holes could potentially survive to the present day, thus contributing to the energy density of dark matter. In this work, we focus on light primordial black holes with masses $M_{\rm PBH} \lesssim 10^{9}~{\rm g}$ that, due to the memory burden effect, are currently evaporating, emitting high-energy particles, among which neutrinos, in the local Universe. Analyzing the latest IceCube data, we place novel constraints on the combined parameter space of primordial black holes and the memory burden effect. We also study the projected reach of future neutrino telescopes such as IceCube-Gen2 and GRAND. We find that the neutrino observations are crucial to probe scenarios with highly-suppressed evaporation and light masses for primordial black holes.
\end{abstract}

\maketitle

{\bf Introduction.  }
The idea that black holes may have formed in the early universe from the direct collapse of overdensities, prior to the formation of any stellar object, has been explored for over 50 years~\cite{Zeldovich:1967lct, Hawking:1971ei, Carr:1974nx}. These intriguing objects --referred to as Primordial Black Holes (PBHs)-- have attracted significant attention over the last decade, especially in the context of Dark Matter (DM) and gravitational wave searches.  

PBHs are metastable objects and emit fundamental particles as they evaporate through Hawking radiation. In the standard scenario, only PBHs with masses above $10^{15}$ grams would evaporate in a timescale longer than the current Hubble time, and could thus be considered viable dark matter candidates. However, strong constraints have been placed, allowing for PBHs as the sole dark matter component only in the asteroid mass range of $10^{17}{\rm g}\lesssim M_{\rm PBH} \lesssim 10^{22}{\rm g}$~\cite{Carr:2016drx, Green:2020jor, Carr:2020gox, Carr:2021bzv}. PBHs with $M_{\rm PBH} \lesssim 10^{15}$ g --which would have entirely evaporated by the time the Universe reaches its current age-- may still play important roles in the production of the baryon asymmetry of the Universe~\cite{Fujita:2014hha, Hamada:2016jnq, Morrison:2018xla, Chen:2019etb, Perez-Gonzalez:2020vnz, Datta:2020bht, Hooper:2020otu, JyotiDas:2021shi, DeLuca:2021oer, Bernal:2022pue, Calabrese:2023key, Calabrese:2023bxz, Schmitz:2023pfy, Barman:2024slw, Gunn:2024xaq}, gravitational wave emission~\cite{Papanikolaou:2020qtd, Domenech:2020ssp, Papanikolaou:2022chm, Ireland:2023avg, Domenech:2024wao}, or dark matter generation~\cite{Bernal:2020kse, Gondolo:2020uqv, Bernal:2020ili, Bernal:2020bjf, Cheek:2021odj, Cheek:2021cfe, Samanta:2021mdm, Bernal:2021yyb, Bernal:2021bbv, Sandick:2021gew, Bernal:2022oha, Cheek:2022mmy, Gehrman:2023qjn, Bertuzzo:2024fns}, depending on their mass.

The semi-classical phenomenon of black hole evaporation relies on the assumption that the black hole remains classical throughout its lifetime~\cite{Hawking:1975vcx}. However, it is increasingly clear that this model may not be self-consistent, indicating the need for new physics, particularly regarding the information loss paradox~\cite{Preskill:1992tc}. Indeed, Hawking’s result ignores the back-reaction of emission on the quantum state of the black hole. This effect becomes however crucial when the energy of the emitted quanta is comparable to the black hole’s total energy. Recent studies~\cite{Dvali:2018xpy, Dvali:2020wft, Dvali:2024hsb} have suggested that the back-reaction may lead to a universal ``memory burden'' effect. This is caused by the fact that the information stored in a system resists its decay, due to the response of the quantum modes associated to the enthropic degrees of freedom. 
Hence, as the black hole’s mass falls below a certain threshold, back-reaction becomes significant, slowing evaporation and potentially extending its lifetime. This implies that black holes below $10^{15}$ grams could be still evaporating by now, thus leading to interesting phenomenological implications~\cite{Balaji:2024hpu, Barman:2024iht, Bhaumik:2024qzd, Barman:2024ufm, Kohri:2024qpd, Jiang:2024aju}. 
Moreover, the memory burden effect also reopens the possibility for light PBHs with masses below $10^9$ grams to be viable dark matter candidates~\cite{Dvali:2020wft, Dvali:2021byy, Alexandre:2024nuo, Thoss:2024hsr, Haque:2024eyh}.

\begin{figure}[t!]
    \centering
    \includegraphics[width=1.0\columnwidth]{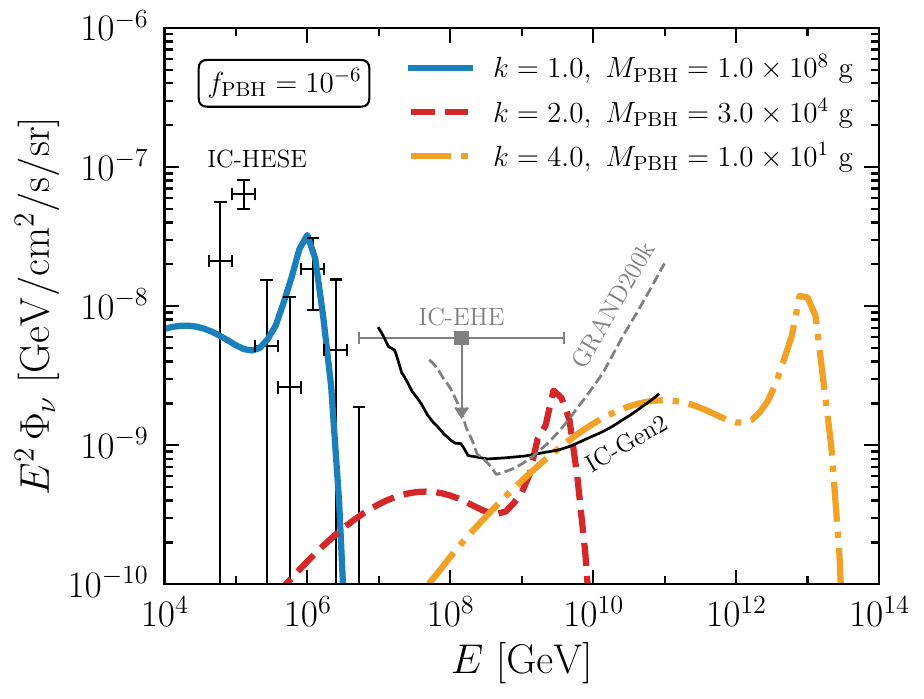}
    \caption{{\bf All-flavor-sum neutrino flux from memory-burdened PBHs.} The different shading lines refer to three benchmark values of the PBH mass $M_{\rm PBH}$ and the memory-burden parameter $k$, with $f_{\rm PBH}=10^{-6}$. Also shown are the 7.5-year IceCube HESE data~\cite{IceCube:2020wum} (black points), the 7-year IceCube EHE upper bound~\cite{IceCube:2016uab} (gray point), and the 3-year flux sensitivity of IceCube-Gen2~\cite{IceCube:2019pna,IceCube-Gen2:2020qha} (thin solid line) and GRAND200k~\cite{GRAND:2018iaj} (thin dashed line).}
    \label{fig:flux}
\end{figure}

\begin{figure*}[t!]
    \centering
    \includegraphics[width=1.0\textwidth]{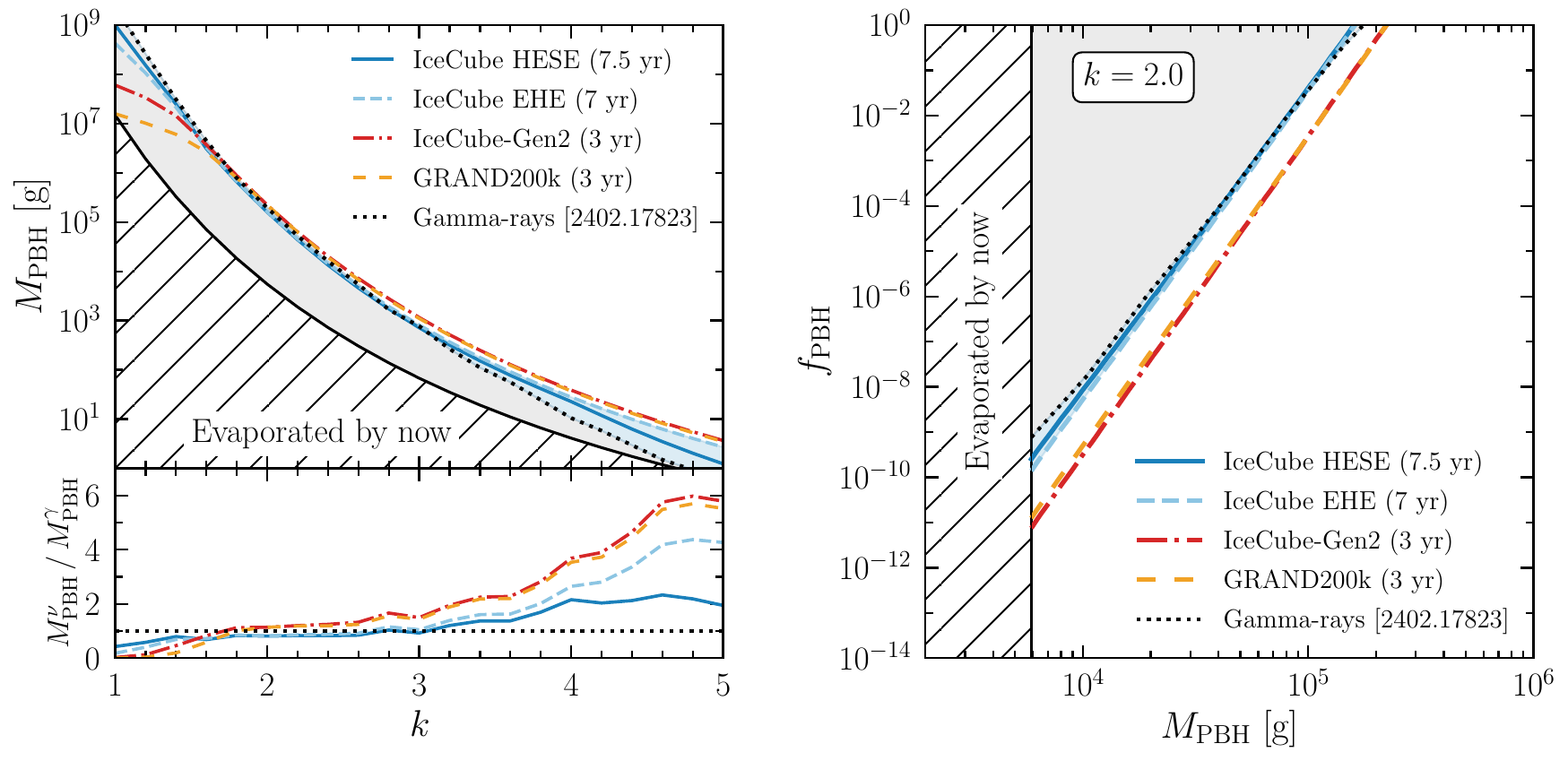}
    \caption{{\bf Neutrino constraints on the parameter space of PBHs and memory burden effect.} The colored lines refer to neutrino limits placed at 95\% C.L. in the $M_{\rm PBH}$-$k$ plane with $f_{\rm PBH} = 1$ (left panel) and in the $M_{\rm PBH}$-$f_{\rm PBH}$ with $k=2.0$ (right panel). In both panels, the white regions correspond to memory-burdened PBHs as viable DM candidates, while the hatched regions to completely-evaporated PBHs. Also shown with dotted black lines are the bounds placed by gamma-ray observations~\cite{Thoss:2024hsr}. In the left panel, the lower plot shows the ratio between the new neutrino limits and the gamma-ray one.}
    \label{fig:limits}
\end{figure*}

In this Letter, we point out that a crucial probe of the existence of memory-burdened PBHs is represented by the observation of high-energy neutrinos as the only unimpeded messenger at high energies. Previous studies have so far explored the steady PBH emission of low-energy neutrinos~\cite{Halzen:1995hu, Dasgupta:2019cae, Lunardini:2019zob, Calabrese:2021zfq, DeRomeri:2021xgy, Bernal:2022swt, Liu:2023cqs}, the final neutrino burst from a nearby PBH~\cite{Capanema:2021hnm, Perez-Gonzalez:2023uoi, DeRomeri:2024zqs}, and the existence of relic neutrinos from fully evaporated light PBHs~\cite{Bugaev:2000bz, Bugaev:2002yt, Wu:2024uxa}, relying on the semi-classical approximation of PBH evaporation. We here address the possibility of detecting the emission of high-energy neutrinos from a population of PBHs with masses $M_{\rm PBH} \lesssim 10^{9}~{\rm g}$ evaporating today thanks to the memory burden effect. We compute the expected neutrino flux at Earth, assuming that these PBHs account for a significant fraction of the dark matter energy density (see Fig.~\ref{fig:flux}). Then, by comparing our predictions with the latest and projected observations from high-energy neutrino telescopes, we establish tighter constraints on the parameter space of PBHs and the associated memory burden effect (see Fig.~\ref{fig:limits}).

{\bf Memory burden effect.  }
It is commonly accepted that PBHs should emit particles due to quantum effects. The emitted radiation has a nearly-thermal spectrum peaked at the so-called Hawking temperature defined as
\begin{equation}
    \label{eq:haw}
    T_{\rm H} = \frac{1}{8 \pi G M_{\rm PBH}} \simeq 10^{4}\left(\frac{10^9~{\rm g}}{M_{\rm PBH}}\right)~{\rm GeV}\,,
\end{equation}
with $G$ being the gravitational constant. The source of this emission is the energy of the PBH gravitational field, so as the radiation is emitted the PBH loses mass at a rate
\begin{equation}
\label{eq:mrate}
    \frac{{\rm d}M_{\rm PBH}}{{\rm d}t} = - \frac{\mathcal{G} \,g_{\rm SM} }{30720 \pi \, G^2 M_{\rm PBH}^2}\,,
\end{equation}
where $\mathcal{G} \sim 3.8$ \cite{Page:1976wx}
is a gray-body factor accounting for the back-scattering of the emitted radiation over the PBH gravitational field, and $g_{\rm SM} \sim 102.6$  \cite{Mazde:2022sdx} is the spin-averaged number of relativistic degrees of freedom in the Standard Model at a temperature $T_{\rm H}$. In the standard picture, the evaporation process continues until the whole PBH mass is converted into radiation. It is then possible to compute the time of complete evaporation as
\begin{equation}
\label{eq:taubh}
    \tau_{\rm PBH} = \frac{10240 \pi G^2 M_{\rm PBH}^3}{\mathcal{G} g_H} \simeq 4.4\times 10^{17}\left(\frac{M_{\rm PBH}}{10^{15}~{\rm g}}\right)^3~{\rm s}\,.
\end{equation}
Therefore, PBHs with a mass $M_{\rm PBH} \lesssim 10^{15} \rm g$ have a lifetime lower than the age of the Universe and would have evaporated by now.

If memory burden is considered, the evaporation process of a PBH can be divided in two phases: an initial semi-classical Hawking-like phase, during which the PBH evolves according to the standard picture, and a second ``burdened phase'', characterized by the stabilization of the PBH by memory burden. At the end of the semi-classical phase, whose duration can be defined as~\cite{Haque:2024eyh}
\begin{equation}
    t_q = \tau_{\rm PBH}(1-q^3)\,,
\end{equation}
the PBH mass results to be a fraction $q$ of its initial mass
\begin{equation}
    M^{\rm mb}_{\rm PBH} = q\,M_{\rm PBH} \,,
\end{equation}
where we take $q = 1/2$ as the memory burden is expected to become relevant, at latest, when the black hole has lost half of its mass. {\color{black}{We note that, in our analysis based on the present PBH evaporation into neutrinos, the choice of $q$ only affects how our results are interpreted in terms of the initial PBH mass.}} For $t \geq t_q$, the quantum effects begin to dominate. The information stored on the event horizon of the PBH back-reacts and slows down the decay rate by a certain negative power of the PBH entropy $S(M_{\rm PBH}) = 4 \pi G M_{\rm PBH}^2$ as
\begin{equation}
\label{eq:mbrate}
    \frac{{\rm d}M^{\rm mb}_{\rm PBH}}{{\rm d}t} = \frac{1}{S(M_{\rm PBH})^{k}}\frac{{\rm d}M_{\rm PBH}}{{\rm d}t} \quad {\rm with} \quad k>0\,.
\end{equation}
Integrating the above equation, we get the time evolution of the PBH mass during the burdened phase:
\begin{equation}
    M_{\rm PBH}^{\rm mb}(t) = M_{\rm PBH}^{\rm mb} \left[ 1- \Gamma_{\rm PBH}^{(k)}(t-t_q)\right]^{1/(3+2k)} \,,
    \label{eq:Mtime}
\end{equation}
where
\begin{equation}
    \Gamma_{\rm PBH}^{(k)} = \frac{\mathcal{G}\,g_{\rm SM}}{7680 \pi}2^k(3+2k)M_P\left(\frac{M_{\rm P}}{M_{\rm PBH}^{\rm mb}}\right)^{3+2k}\,,
\end{equation}
with $M_P=(8 \pi G)^{-1/2}$ being the reduced Planck mass. The total evaporation time is therefore equal to
\begin{equation}
    \tau_{\rm PBH}^{(k)} = t_q + (\Gamma_{\rm PBH}^{(k)})^{-1} \simeq (\Gamma_{\rm PBH}^{(k)})^{-1}\,.
\end{equation}
This can be several orders of magnitude higher than the standard evaporation time for $k>0$, so the memory burden effect allows for much lighter PBHs to survive until today and contribute to the present DM energy density.

{\bf The high-energy neutrino flux.  }
We here compute the flux of high-energy neutrinos emitted by a population of memory-burdened PBHs which can provide a fraction $f_{\rm PBH}=\Omega_{\rm PBH}/\Omega_{\rm DM}$ of the total DM component of the Universe. For the sake of concreteness, we assume here a monochromatic mass spectrum for the PBH population defined by a mass in the range $10^{-1}~{\rm g} \leq M_{\rm PBH} \leq 10^9 ~{\rm g}$.

The semi-classical neutrino emission rate for a chargeless and non-rotating PBH with mass $M_{\rm PBH}$ is
\begin{equation}
    \frac{{\rm d}^2N_{\nu}}{{\rm d}E{\rm d}t} = \frac{g_{\nu}}{2 \pi} \frac{\mathcal{F}(E,M_{\rm PBH})}{e^{E/T_{\rm H}} + 1} \,,
\end{equation}
with $g_{\nu} = 6$ being the number of internal degrees of freedom of the three neutrino families, and $\mathcal{F}(E,M_{\rm PBH})$ the gray-body factor. In the present analysis, we numerically compute the semi-classical neutrino emission rate by using the code \texttt{BlackHawk}~\cite{Arbey:2019mbc,Arbey:2021mbl} which also includes the secondary neutrino emission according to \texttt{HDMSpectra} hadronization~\cite{Bauer:2020jay}.

When the PBH enters its burdened phase, its emission rate becomes suppressed according to Eq.~\eqref{eq:mbrate}, becoming:
\begin{equation}
    \frac{{\rm d}^2N^{\rm mb}_{\nu}}{{\rm d}E{\rm d}t}  = S(M_{\rm PBH})^{-k} \, \frac{{\rm d}^2N_{\nu}}{{\rm d}E{\rm d}t} \,.
\end{equation}
This spectrum is peaked at the same temperature $T_{H}$ as in the semi-classical phase. For this reason, the energy of the neutrinos emitted today by ``burdened survivor'' PBHs with $M_{\rm PBH}\lesssim$ $10^{9}~{\rm g}$ is very high (see Eq.~\eqref{eq:haw}).

We therefore expect an all-flavour-sum neutrino flux at high energies coming from both the DM halo of our galaxy, and from the extragalactic isotropic DM distribution as
\begin{equation}
    \Phi_\nu = \sum_\alpha\left(\frac{{\rm d}^2\phi^{\rm gal}_{\nu_\alpha}}{{\rm d}E{\rm d}\Omega}+\frac{{\rm d}^2\phi^{\rm egal}_{\nu_\alpha}}{{\rm d}E{\rm d}\Omega}\right)\,,
    \label{eq:flux_tot}
\end{equation}
where $\alpha$ runs over the three neutrino flavors. 
The galactic component, averaged over the whole solid angle, takes the expression
\begin{equation}
    \frac{{\rm d}^2\phi^{\rm gal}_{\nu_\alpha}}{{\rm d}E{\rm d}\Omega} = \frac{f_{\rm PBH}\,\mathcal{J}}{4 \pi M^{\rm mb}_{\rm PBH}} \frac{{\rm d}^2 N^{\rm mb}_{\nu_\alpha}}{{\rm d}E{\rm d}t}\,,
    \label{eq:gal}
\end{equation}
where the normalization of the flux and the neutrino emission rate are fixed by the today PBH mass $M_{\rm PBH}^{\rm mb}$, and $\mathcal{J} = 2.22 \times 10^{22}~{\rm GeV/cm^2/sr}$ is the averaged J-factor. We consider a NFW density profile with scale radius of 25 kpc and a local DM density of $0.4~{\rm GeV/cm^3}$~\cite{Iocco:2015xga,Benito:2019ngh,Benito:2020lgu}.
The extragalactic component is given by
\begin{equation}
\frac{{\rm d}^2 \phi^{\rm egal}_{\nu_\alpha}}{{\rm d}E {\rm d}\Omega} = \frac{f_{\rm PBH} \,\rho_{\rm DM}}{4 \pi M^{\rm mb}_{\rm PBH}} \int_{t_{\rm min}}^{t_{\rm max}} {\rm d}t \,\left[1+z(t)\right] \frac{{\rm d}^2N^{\rm mb}_{\nu_\alpha}}{{\rm d}E{\rm d}t}\,,
\label{eq:egal}
\end{equation}
where $\rho_{\rm DM} = 1.26\times 10^{-6}~{\rm GeV/cm^3}$~\cite{Planck:2018vyg}, the neutrino emission rate is computed by taking into account the effect of redshift $z(t)$ on the energy and the time evolution of the PBH mass as given by Eq.~\eqref{eq:Mtime}, and the time integral is performed from the time of matter-radiation equality ($t_{\rm min}$) to the age of the Universe ($t_{\rm max}$).

Fig.~\ref{fig:flux} shows the all-flavor-sum neutrino flux $\Phi_\nu$ computed for three benchmark choices of the parameters $k$ and $M_{\rm PBH}$, taking $f_{\rm PBH} = 10^{-6}$. As expected, the lighter the PBH mass, the higher the energies of the neutrinos emitted via evaporation. In all the cases, we find that the neutrino flux is dominated by the galactic component, with the primary (secondary) emission manly contributing to the peak (tail). On the other hand, the extragalactic component emission is almost negligible. Similar neutrino fluxes may also arise from the mergers of galactic memory-burdened PBHs that have resumed Hawking evaporation~\cite{Zantedeschi:2024ram}. In the plot, we also show the current IceCube observations given by the 7.5-year High-Energy Starting Events (HESE) data sample~\cite{IceCube:2020wum} (black data points on the left) and the 7-year Extremely High-Energy (EHE) upper bound $E^2 \Phi_\nu \leq 5.9 \times 10^{-9}~{\rm GeV/cm^2/s/sr}$ from $1.0 \times 10^6$ to $4.0 \times 10^9$~GeV at 90\% C.L.~\cite{IceCube:2016uab} (gray data point in the middle), as well as the 3-year future sensitivity of IceCube-Gen2~\cite{IceCube:2019pna,IceCube-Gen2:2020qha} (thin solid black line) and GRAND200k~\cite{GRAND:2018iaj} (thin dashed gray line) telescopes. We note that the scenario of memory-burden PBHs could be also probed with the currently-operating KM3NeT telescope with sensitivity to neutrinos up to few tens of PeV~\cite{KM3Net:2016zxf, KM3NeT:2024paj}.

{\bf Statistical analysis.  }
We analyze the latest data collected by IceCube to derive conservative limits on the memory-burdened PBH parameter space. We employ a background-agnostic likelihood analysis based on the following likelihood function
\begin{equation}
    \mathcal{L}\left(f_{\rm PBH}; M_{\rm PBH}, k \right) = \prod_{i}^{n_{\rm data}}\left\{\begin{array}{l l}
    \mathcal{P}\left(d_i|\mu_i\right) & \mu_i > d_i \\
    1 &\mu_i \leq d_i 
    \end{array}\right. \,,
\end{equation}
where $\mathcal{P}$ is the probability distribution function of the data $d_i$ with expected mean $\mu_i\left(f_{\rm PBH}; M_{\rm PBH}, k \right)$, and the index $i$ runs over the number of data $n_{\rm data}$. We hence compute the upper bound on $f_{\rm PBH}$ for each choice of $M_{\rm PBH}$ and $k$ parameters by taking $\Delta \chi^2 = -2\ln \mathcal{L}$ and assuming the Wilks theorem with one degree of freedom.

In case of the 7.5-year IceCube HESE data~\cite{IceCube:2020wum}, we consider the frequentist flux measurements as data and, consequently, the probability distribution $\mathcal{P}$ to be a Gaussian distribution. On the other hand, in case of the 7-year IceCube EHE data~\cite{IceCube:2016uab}, we take a Poisson distribution with a single detected event in the entire energy range $\left[E_{\rm min},E_{\rm max}\right]$ of the experiment and the expected number of events from PBHs given by
\begin{equation}
    n^{\rm PBH}_{\rm events} = 4 \pi\, T_{\rm obs} \int_{E_{\rm min}}^{E_{\rm max}} {\rm d}E\,\Phi_\nu(E)\,A_{\rm eff}(E)\,,
\end{equation}
with $A_{\rm eff}(E)$ being the IceCube EHE effective area, and $T_{\rm obs}=7~{\rm yr}$ the data-taking time.

For the forecast analysis of the future telescopes IceCube-Gen2~\cite{IceCube:2019pna,IceCube-Gen2:2020qha} and GRAND200k~\cite{GRAND:2018iaj}, we instead employ a different method. We assume that no events will be observed after 3 years of data-taking in the whole energy range and compute the upper limit on $f_{\rm PBH}$ at 95\% C.L. by taking $n^{\rm PBH}_{\rm events} < 3.09$ according to the approach of Feldman and Cousins with zero background~\cite{Feldman:1997qc}. In this case, we estimate the effective area of the two detectors by their flux sensitivity following Ref.~\cite{Chianese:2021htv}. We emphasize that the assumption of zero detected events is not conservative bur rather realistic. Indeed, the cosmogenic neutrino flux, which is among the primary targets of future neutrino telescopes, could be well below the experimental sensitivity~\cite{Ehlert:2023btz,Berat:2024rvf}. Moreover, the emission of ultra-energy neutrinos from different classes of sources is also affected by large uncertainties~\cite{Biehl:2017hnb,Rodrigues:2020pli,Condorelli:2022vfa}.

Therefore, in the present analysis we take into account the spectral distribution only in case of the IceCube HESE data sample. For the other cases, our results are based on the integrated number of events in the entire energy range where each telescope is sensitive. We do not analyse the angular distribution of the PBH neutrino flux which is expected to be greater toward the center of our galaxy according to the DM halo distribution. For these reasons, our results are conservative and future analysis may actually strengthen our results and tighten the constraints.

{\bf Results.  }
We report the main result of our analysis in Fig.~\ref{fig:limits}. In the left panel, the different lines represent the constraints on the PBH mass $M_{\rm PBH}$ as a function of the memory-burden parameter $k$ assuming $f_{\rm PBH} = 1$. This means that the white region corresponds to scenarios where PBHs can account for the total of the DM component of the Universe. The hatched region marks the parameter space where PBHs have completely evaporated in cosmological times. The colored lines show our new constraints placed with current neutrino data (continuous and dashed lines) and future ones (dot-- and long-- dashed lines), while the dotted black line refers to the previous gamma-ray limits~\cite{Thoss:2024hsr}. Remarkably, we find that the conservative neutrino limits are tighter than the gamma-ray ones (dotted black line) for large values of $k$, as highlighted in the lower panel showing the ratio among the limits. For instance, in case of $k=5.0$, the present (/future) neutrino limits are stronger by a factor of 2 (/6) than gamma-ray ones. This result owes indeed to the fact that larger values for $k$ correspond to lighter still-survived PBHs and, consequently, to PBH particle emission at higher energies where gamma-rays are mainly absorbed.

In the right panel of Fig.~\ref{fig:limits}, we show the constraints projected in the plane $M_{\rm PBH}$-$f_{\rm PBH}$ for the specific case of $k = 2.0$. For $f_{\rm PBH}=1$, the neutrino and gamma-ray limits are similar, implying $M_{\rm PBH} \gtrsim 2 \times 10^5~{\rm g}$ for PBHs as viable DM candidates. On the other hand, we find that, for light PBHs near the evaporation threshold (hatched region), neutrino observations improve the constraints on $f_{\rm PBH}$ up to two orders of magnitude. 

{\bf Conclusions.  }
In the present Letter, we address a scenario in which PBHs with masses $M_{\rm PBH} \lesssim 10^{9}~{\rm g}$ evaporate today owing to the ``memory burden'' effect. The flux of high-energy neutrinos emitted in this process should be in principle observed, and the mechanism can therefore be used to constrain the PBH and memory burden effect parameter space.

We estimate the expected flux and perform an analysis of currently available high energy neutrino data, presenting novel constraints in the PBH parameter space. Even with a very conservative statistical analysis, our constraints are highly competitive with respect to the existing ones, which are obtained mainly by gamma-ray observations. We also perform a forecast analysis to determine the constraints that can be placed with upcoming neutrino telescopes.

In summary, our analysis demonstrates that high-energy neutrinos represent a powerful tool to probe the memory-burden effect of light PBHs. We expect that the employment of more tailored statistical methods will already bring improvements in the constraints even before the results of upcoming neutrino experiments.

{\bf Acknowledgements.  }
We acknowledge the support by the research project TAsP (Theoretical Astroparticle Physics) funded by the Istituto Nazionale di Fisica Nucleare (INFN). The work of FI and NS is further supported by the research grant number 2022E2J4RK ``PANTHEON: Perspectives in Astroparticle and Neutrino THEory with Old and New messengers'' under the program PRIN 2022 funded by the Italian Ministero dell’Università e della Ricerca (MUR).

\bibliography{bibliography}
\end{document}